\documentclass[aps,prl,preprint,onecolumn,superscriptaddress]{revtex4-1}
\usepackage{amsmath,amssymb}
\usepackage{graphicx}
\usepackage{float}
\usepackage{natbib}
\usepackage{color}
\usepackage[colorlinks=true,bookmarks=false,citecolor=blue]{hyperref}
\usepackage{lmodern}
\usepackage[version=4]{mhchem}
\usepackage[most]{tcolorbox}
\usepackage{tikz}
\usepackage{varwidth}
\usepackage{wrapfig}
\usepackage{capt-of}
\usepackage{paracol}
\usepackage{threeparttable}

\begin{document}

\title{Nanostructured transition metal dichalcogenide multilayers for advanced nanophotonics}

\author{Battulga Munkhbat}
\affiliation{Department of Physics, Chalmers University of Technology, 412 96, Göteborg, Sweden}

\author{Betül Küçüköz}
\affiliation{Department of Physics, Chalmers University of Technology, 412 96, Göteborg, Sweden}

\author{Denis G. Baranov}
\affiliation{Department of Physics, Chalmers University of Technology, 412 96, Göteborg, Sweden}
\affiliation{Center for Photonics and 2D Materials, Moscow Institute of Physics and Technology, Dolgoprudny, 141700, Russia}

\author{Tomasz J. Antosiewicz}
\affiliation{Faculty of Physics, University of Warsaw, Pasteura 5, 02-093 Warsaw, Poland.}

\author{Timur O. Shegai}
\email[]{timurs@chalmers.se}
\affiliation{Department of Physics, Chalmers University of Technology, 412 96, Göteborg, Sweden}

\begin{abstract}
Transition metal dichalcogenides (TMDs) attract significant attention due to their exceptional optical, excitonic, mechanical, and electronic properties. Nanostructured multilayer TMDs were recently shown to be highly promising for nanophotonic applications, as motivated by their exceptionally high refractive indexes and optical anisotropy. Here, we extend this vision to more sophisticated structures, such as periodic arrays of nanodisks and nanoholes, as well as proof-of-concept waveguides and resonators. We specifically focus on various advanced nanofabrication strategies, including careful selection of resists for electron beam lithography and etching methods. The specific materials studied here include semiconducting WS$_2$, in-plane anisotropic ReS$_2$, and metallic TaSe$_2$, TaS$_2$ and NbSe$_2$. The resulting nanostructures can potentially impact several nanophotonic and optoelectronic areas, including high-index nanophotonics, plasmonics and on-chip optical circuits. The knowledge of TMD material-dependent nanofabrication parameters developed here will help broaden the scope of future applications of these materials in all-TMD nanophotonics.

\end{abstract}

\maketitle

\section{Introduction}

Nanostructuring is one of the most attractive and facile ways to enrich material properties. In layered transition metal dichalcogenide (TMD) materials, nanostructuring can be readily introduced through exfoliation. This approach is now routinely exploited down to the monolayer limit and the resulting TMD monolayers exhibit remarkable physical and chemical properties. For instance, the recent interest in two-dimensional (2D) semiconductors is largely motivated by the discovery of a direct bandgap in monolayer MoS$_2$ \cite{splendiani2010emerging, mak2010atomically}. Furthermore, TMD monolayers can be assembled into van der Waals (vdW) heterostructures, which enrich material properties through formation of Moir\'e superlattices \cite{geim2013van}. The possible confinement of TMD materials, however, is not limited to the out-of-plane direction and thus to monolayer physics. A straightforward alternative is in-plane confinement, which can be achieved through various kinds of structural etching and can be realized in both TMD mono- and multilayers \cite{verre2019transition, busschaert2020transition, zhang2019guiding, green2020optical, munkhbat2020transition}. Such nanostructuring can introduce an additional degree of freedom in tailoring physical and chemical properties of TMD materials on various scales, including edge physics and nanophotonics.

Applying the nanopatterning approach to TMD multilayers can bear interest since multilayers possess several attractive optical and electronic properties, which can be potentially enhanced and/or enriched through nanofabrication. 
Optical properties of TMDs are determined by their atomic lattice and the corresponding electron band structure.
Many TMDs feature exitonic resonances in the visible range that possess high oscillator strength and are stable both in mono- and multi-layer materials at room temperature. 
Owing to Kramers-Kronig relations, these exitonic transitions endow TMDs with relatively high refractive indices ($n\sim$ 4) at energies just below the respective transition \cite{li2014measurement, verre2019transition, ermolaev2020broadband}. 
At even lower energies (longer wavelengths) where the refractive index remains large but the material's absorption coefficient quickly vanishes, these materials can be exploited for lossless waveguiding and light confinement \cite{hu2017probing, hu2017imaging, babicheva2018near, ermolaev2021giant, ermolaev2020broadband}.
Furthermore, all TMDs are naturally anisotropic due to their vdW nature, endowing them with an inherent optical uniaxial permittivity with $\varepsilon_{xx}=\varepsilon_{yy}\neq\varepsilon_{zz}$ \cite{hu2017probing, green2020optical, ermolaev2021giant}.
In some materials a distortion of the planar hexagonal lattice, such as in \ce{ReS2}, leads to biaxial anisotropy with different values of the in-plane permittivity tensor elements, i.e. $\varepsilon_{xx}\neq\varepsilon_{yy}$ and both being different from the out-of-plane $\varepsilon_{zz}$ \cite{gogna2020self,wang2021prediction,ma2021ghost}. Nonlinear optical properties of multilayer TMDs, such as second-harmonic generation (SHG), also attract research attention \cite{busschaert2020transition,nauman2021tunable,tselikov2021double}. Here, bulk materials characterised by broken inversion symmetry, such as multilayer ReS$_2$ or hBN, play an important role \cite{Zhao2018,caldwell2019photonics}.

Nanostructured multilayer TMDs are thus promising candidates for future nano- and optoelectronic applications \cite{wang2018colloquium}. Although several examples of TMD nanostructuring have already been reported in the literature \cite{verre2019transition, busschaert2020transition, zhang2019guiding, green2020optical, roxlo1987edge, wang2020exciton, nauman2021tunable, tselikov2021double}, detailed studies of nanopatterned TMDs are just starting to appear. Thus, a material-oriented study focusing on advanced nanofabrication procedures for TMD multilayers and demonstrating a potential of all-TMD nanophotonics is of high contemporary interest.


In this paper, we present fabrication of TMD nanostructures confined in one, two, or all three dimensions, and experimentally measure and analyze some of their optical properties both in the linear and non-linear regimes. 
Using bulk WS$_2$, ReS$_2$, and TaSe$_2$ crystals as a starting point, we fabricated nanodisk/nanohole arrays, waveguides, microring cavities, photonic crystal cavities, as well as individual nanodisks and nanorods.
In the linear regime, nanopatterned WS$_2$ arrays exhibit various resonant features including Mie-type resonances and diffractive modes.
In the non-linear regime, TMD arrays exhibit selective enhancement of the second-harmonic generation. Nanopatterned WS$_2$ structures also hold promise for optical waveguiding in e.g. bus-coupled microring structures, and for light localization in photonic crystal cavities as demonstrated herein using numerical modelling. Nanopatterning is feasible also for in-plane anisotropic TMDs such as \ce{ReS2} as well as uniaxial metallic TMDs represented here by TaSe$_2$ and NbSe$_2$. An example of the former includes hexagonal nanodisk arrays, while for the latter we demonstrated accurate fabrication of nanoantennas based on single elements as well as dimers. The results reported here, along with mechanical flexibility of TMDs \cite{chang2013high}, broaden the scope of potential applications of TMD materials in nanophotonics and (opto)electronics.



\begin{figure}
\includegraphics[width=1\textwidth]{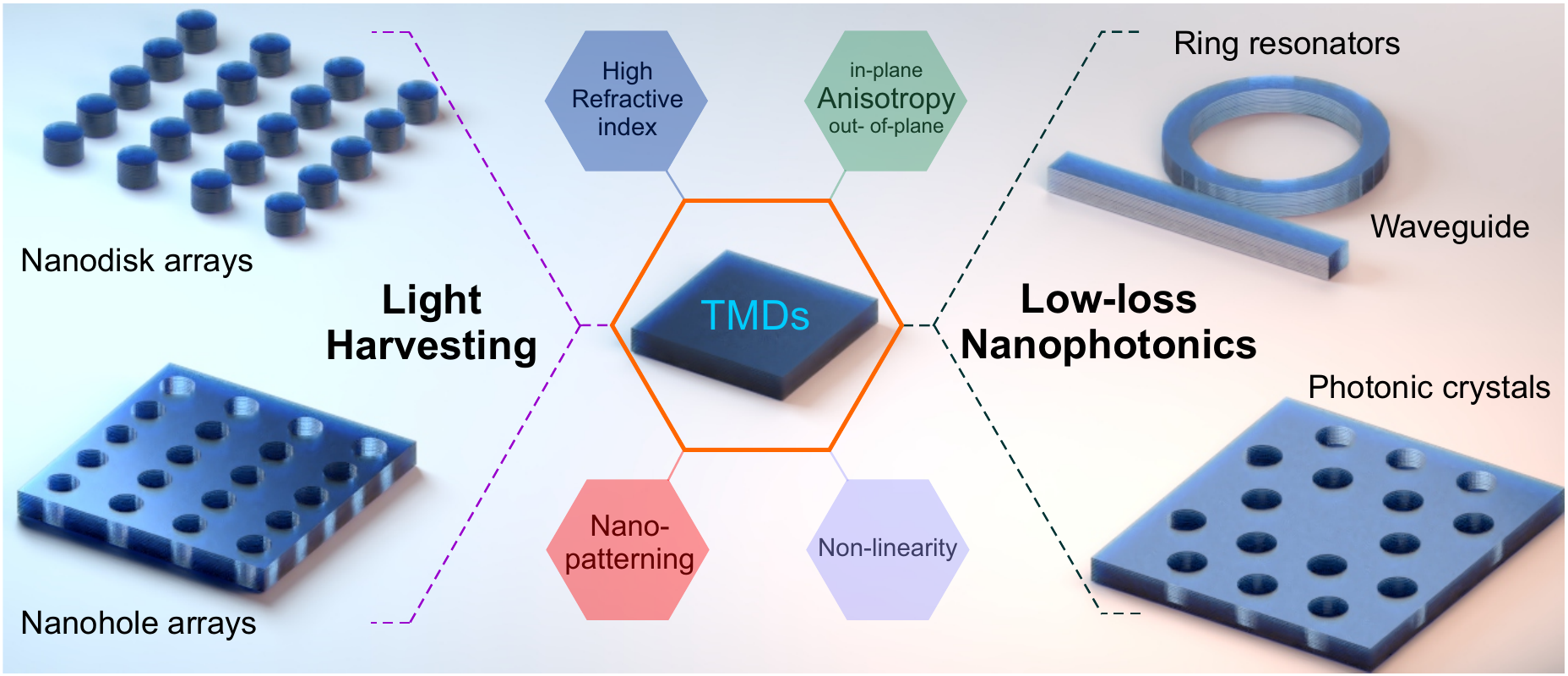}
\caption{\textbf{Roadmap for TMD nanophotonics.} Exploring nanostructured TMDs for future applications in \textit{(left)} light absorption, and \textit{(right)} low-loss nanophotonics. This vision is based on remarkable properties of multilayer TMDs, such as high refractive index, optical anisotropy, and optical nonlinearity. The purpose of this work is to demonstrate improvement in optical response of TMDs through nanostructuring.}
\label{Fig1}
\end{figure}

\autoref{Fig1} conceptualizes the approach to TMD nanophotonics that we follow in this study. TMD materials combine a spectral region of high absorption associated with their excitonic resonances, and a low optical loss region at energies below the A-exciton resonance. These two regimes determine two different application areas where nanostructured TMDs can be exploited: (i) light absorption, and (ii) light confinement and guiding, correspondingly.
Accordingly, we split our investigation of nanostructured TMDs into two parts. In the first, we demonstrate the capabilities of nanostructured TMDs for light harvesting purposes by utilizing the absorptive behavior of TMDs in the visible frequency range. In the second part, we present and discuss the capabilities of nanostructured TMDs for light confinement and guiding in the long-wavelength part of the spectrum in which these materials exhibit low losses.
In the following, we examine different types of nanostructures suited for both objectives presented in this study, discuss methods with which to fabricate them \cite{verre2019transition, green2020optical, munkhbat2020transition}, characterize their optical response using various spectroscopic methods, and provide numerical simulation results. 

\section{TMD nanostructures for light absorption}

We begin our discussion by examining periodic arrays of TMD nanodisks and nanoholes, whose usefulness is based on the high refractive index of TMDs \cite{verre2019transition, li2014measurement}. These systems possess optical resonances associated with the modes of individual elements, such as Mie resonances of individual disks, and lattice modes formed due to far-field couplings in periodic systems \cite{verre2019transition, munkhbat2020transition, babicheva2018near, de2007colloquium, babicheva2019lattice}.


\subsection{Nanodisk arrays}

\begin{figure}
\centering
\includegraphics[width=0.9\textwidth]{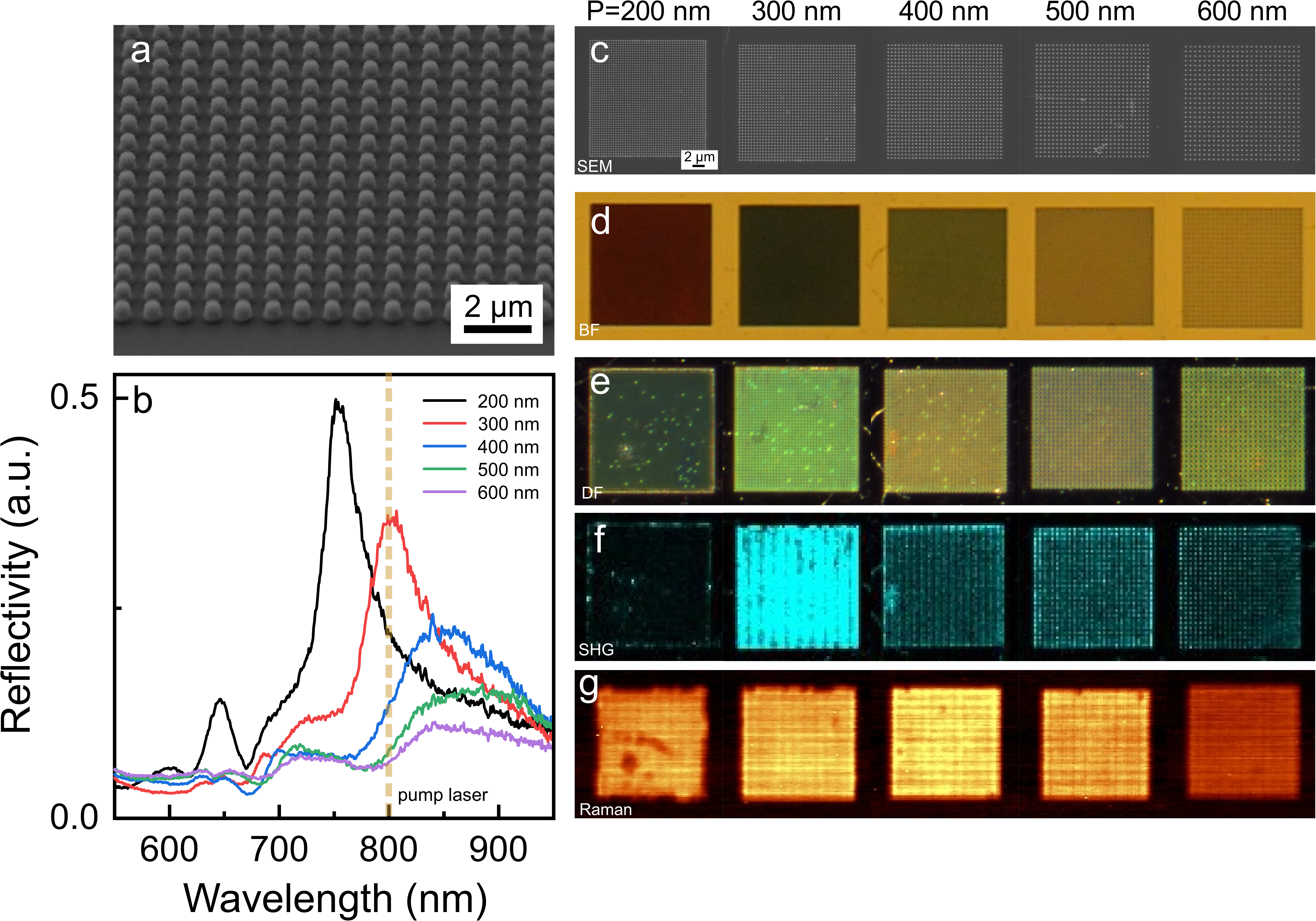}
\caption{\textbf{Arrays of TMD nanodisks.} (a) Tilted-view SEM image of TMD nanodisk array with diameter $D=250$~nm and pitch $P=200$ nm. (b) Reflection spectra from TMD nanodisk arrays with $D=250$~nm and $P$ varying from 200 to 600 nm. (c-g) Images of TMD nanodisk arrays with $D=250$~nm and $P$ varying from 200 to 600 nm using: (c) scanning electron microscopy, (d) bright-field optical microscopy, (e) dark-field optical microscopy, (f) second-harmonic generation microscopy, (g) Raman microscopy.}
\label{Fig2}
\end{figure}

\autoref{Fig2} summarizes our results on nanodisk arrays made of multilayer WS$_2$. The samples were fabricated by a combination of electron beam lithography (EBL) and reactive ion etching (RIE), see Methods. For simplicity, in this study we restrict ourselves to square nanodisk arrays. However, more complex array symmetries are equally possible with the use of EBL. As a proof-of-concept, first, we fabricated WS$_2$ nanodisk lattices with various pitches ($P$: in the 200-600 nm range with a step of 100 nm) and disk diameters ($D$: 200 and 250 nm) on a transparent SiO$_2$ substrate (\autoref{Fig2}a). Here, we take advantage of our previous nanofabrication recipes \cite{verre2019transition, green2020optical} and develop them further. 

Briefly, mechanically exfoliated WS$_2$ flakes were transferred onto a SiO$_2$ substrate using a dry-transfer method \cite{Castellanos2014transfer}.  Nanofabrication of TMD nanostructures on silicon substrates with thin thermal oxides is relatively straightforward and was addressed in our previous studies \cite{verre2019transition,green2020optical,munkhbat2020transition}. On the other hand, it is demanding to fabricate TMD nanostructures on transparent, low refractive index, and non-conductive substrates, such as SiO$_2$. Nanostructures on microscope cover glasses are useful for practical applications, as they allow to monitor the sample using high numerical aperture (NA) immersion objectives in both transmission and reflection modes. Specifically, use of a negative resist on top of a non-conductive glass substrate, a method which yields high-quality TMD nanostructures, increases fabrication complexity. Therefore, one of the practical goals of this study is to develop a recipe for fabrication of high-quality TMD nanostructures not only on standard silicon substrates, but also on other substrates such as non-conductive, transparent thin glass substrates for future nanophotonic applications. 

Our recipe to fabricate TMD nanodisk arrays on a glass substrate with a negative resist is the following. We start with deposition of a negative e-beam resist (MaN-2405) on the substrate by spin coating at 3000 rpm and followed by soft baking at 90$^{\circ}$C to prepare an etching mask. It is worth mentioning that other standard negative resists e.g. HSQ can also be used to prepare the etching mask. The main issue for applying the resist homogeneously onto TMD flakes on a substrate is adhesion of the resist. Our solution to the problem is applying the TI:Prime adhesion promoter prior to the resist deposition. As an alternative, a standard HMDS adhesion promoter can be used. However, application of HMDS results in less reproducible results since this material is more sensitive to ambient conditions, in contrast to TI:Prime, which is why we chose to work with the latter.

To achieve high-quality densely packed TMD nanodisk arrays with EBL, the resist should be typically pre-coated with a reflective and conductive thin layer of chromium (Cr, 20 nm) in order to (i) suppress electron beam scattering in the resist due to the proximity effect, and (ii) for better focusing of the e-beam on the resist surface. However, removing the Cr layer after the exposure is an issue and is not compatible with the MaN-2405 negative resist, since the standard wet-etchant for Cr chemically modifies the resist during etching, which in turn results in a problem for further development of MaN-2405. One solution to this problem could be replacing the Cr layer with a water-soluble conductive polymer (e-spacer) film. However, using an e-spacer polymer in combination with a thick negative resist on a non-conductive substrate results in poor mask quality for later fabrication stages. Therefore, our solution to the problem is to combine the water-soluble e-spacer with the reflective Cr layer to achieve the best quality sample. To that end, the water-soluble e-spacer is spin-coated directly on the resist at 2000 rpm for 20 s before depositing the Cr layer. This results in a higher quality of the etching mask after the development process, since removal of the Cr layer can be done simply by rinsing with running water for a few seconds. In this process, one needs to be careful and treat the sample gently since the TMD flake is at risk of delamination from the substrate because of weak vdW forces acting between them. After removing the Cr layer, the exposed sample is developed using the MaD-525 developer, followed by rinsing with water. The sample is then dry-etched in an inductively coupled plasma reactive ion etcher (RIE) using CHF$_3$ gas (see Methods). 

Since we use CHF$_3$ as a main etching gas, the top resist may get hardened because of carbon re-deposition in the presence of fluorine ions. This leads to two consequences: (i) a positive one: it allows us to perform longer etching due to the hardened resist mask and (ii) a negative one: it may lead to an unwanted issue of removing the left-over resist after the etching. To resolve the negative consequence of carbon re-deposition, we perform additional oxygen plasma stripping with O$_2$ (40 sccm) after the dry etching. This removes the top hardened resist layer. Subsequently, the left-over resist can be easily removed by exposure to hot acetone (50$^{\circ}$C) for 3 min, followed by rinsing with isopropyl alcohol and deionized water.

After fabrication we characterized the WS$_2$ nanodisk samples using several spectroscopy and microscopy techniques. \autoref{Fig2}b shows normal-incidence reflection spectra from WS$_2$ nanodisk lattices with diameter $D=250$ nm and pitch ($P$) varying from 200 to 600 nm. Their corresponding scanning electron microscopy (SEM), bright-field (BF), and dark-field (DF) images are shown in \autoref{Fig2}c-e, respectively. Our results demonstrate the appearance of lattice modes in WS$_2$ nanodisk arrays that can be tuned by controlling the geometrical parameters of the sample, such as $D$ and $P$. These structures exhibit bright colors across the entire visible range. The colors depend on the structure of the nanoparticle array, including individual nanodisk height and radius, as well as the array arrangement (see \autoref{Fig2}d,e). The peaks in the reflection spectra plotted in \autoref{Fig2}b originate from nanodisk ($D=250$ nm) arrays with various $P$ red-shifts, when $P$ increases from 200 nm to 600 nm. This resonance feature can be attributed to the geometrical lattice mode, since it is known that periodically arranged high-index dielectric nanodisks can exhibit geometrical lattice modes \cite{babicheva2019lattice}. The data obtained from the nanodisk arrays with $D=200$ nm are summarized in Supplementary Figure S1.

Furthermore, we performed additional measurements to demonstrate the angular dispersion behavior of the lattice modes (see Supplementary Figure S2). Our results show that the peaks in the reflection spectra exhibit clear lattice-like angular dispersion behavior. Due to the fact that the peaks in reflection show a clear $P$-dependence with fixed $D$ as well as angular dispersion, they can be interpreted as lattice modes induced by the periodic arrangement of WS$_2$ nanodisks.

Next, we studied second-harmonic generation (SHG) from the fabricated WS$_2$ nanodisk arrays. SHG measurements were carried out using a home-built optical setup, coupled to a tunable (690 nm - 1040 nm) Ti:Sapphire femtosecond laser. Excitation wavelengths were varied between $\lambda_{exc}=770$, 800, 850, 900, and 950 nm to match the resonances of the lattice modes in the WS$_2$ nanodisk arrays with different $P$. 

First, we carried out the power-dependent SHG measurements from an array and obtained the slope-2 in logarithmic scale to prove that the collected signal originates from SHG (Supplementary Figure S3). It is known that unpatterned multilayer WS$_2$ flakes exhibit a negligible SHG signal due to the centrosymmetric nature of the AB-stacked bulk crystal. Therefore, it is interesting to verify whether nanopatterning can enhance and modify SHG in a multilayer WS$_2$ sample. Several examples for SHG imaging ($\lambda_{SHG}=400$ nm) from WS$_2$ nanodisk ($D=250$ nm) arrays with various $P$ under $\lambda_{exc}=800$ nm excitation are shown in \autoref{Fig2}f (note false color). Interestingly, we observed that when the lattices are resonantly pumped with $\lambda_{exc}$ that spectrally overlaps with the lattice modes, SHG imaging shows the highest intensity from the resonant array. One can clearly see that the nanodisk ($D=250$ nm) array with $P=300$ nm shows the highest SHG intensity due to the resonant excitation of $\lambda_{exc}=800$ nm, whereas $P=200$ nm gives much lower SHG intensity because of the poor spectral overlap with the lattice mode. On the other hand, when we pumped all nanodisk arrays by $\lambda_{exc}=770$ nm, the $D=250$ nm array with $P=200$ nm resulted in the highest SHG intensity due to the better resonant excitation (see Supplementary Figures S4 and S6).
The data from $D=200$ nm samples are summarized in Supplementary Figure S5. Our observations suggest that the SHG signal from the WS$_2$ nanodisk arrays can be enhanced by controlling the optical resonances through nanopatterning. Such enhancement occurs despite the absence of broken inversion symmetry in the bulk WS$_2$ crystal. Thus, the origin of SHG in this case is likely the surface and edges exposed through nanopatterning.

This suggests that in order to attain the highest SHG intensity enhancement one should maximize the surface area of the nanostructure and at the same time maintain high electric field intensity at the surface, which calls for a non-trivial optimization procedure. Resonant modes of individual nanoantennas, as well as the diffractive interactions within the array, have been previously shown to enhance the SHG intensity in TMD metasurfaces \cite{busschaert2020transition,nauman2021tunable,tselikov2021double}. The results observed with our arrays can probably be explained by similar mechanisms; given the dimensions of the individual nanodisk and the excitation wavelength, the dipole and quadrupole Mie resonances are expected to play the key role in the SHG enhancement.


To verify that the WS$_2$ flake was etched completely into nanodisk arrays, we performed confocal Raman mapping. The Raman spectra show two clear Raman peaks at 349 cm$^{-1}$ and 415 cm$^{-1}$, revealing that nanopatterned WS$_2$ was not damaged (see Supplementary Figure S7). \autoref{Fig2}g shows confocal Raman maps at 415 cm$^{-1}$ obtained from the WS$_2$ nanodisk arrays. Raman maps confirm the complete removal of WS$_2$ in all regions except for the lattice. It is worth mentioning that Raman mapping was carried out after all other optical characterizations, to prevent potential photodamage by strong laser irradiation. This precaution was motivated by our previous observations of unpatterned WS$_2$ flakes being damaged during Raman mapping, probably because of the spectral overlap of the excitation laser ($\lambda_{exc}=532$ nm) with the absorption band of WS$_2$ that can lead to unwanted heating. 
 
\begin{figure}
\centering
\includegraphics[width=0.8\textwidth]{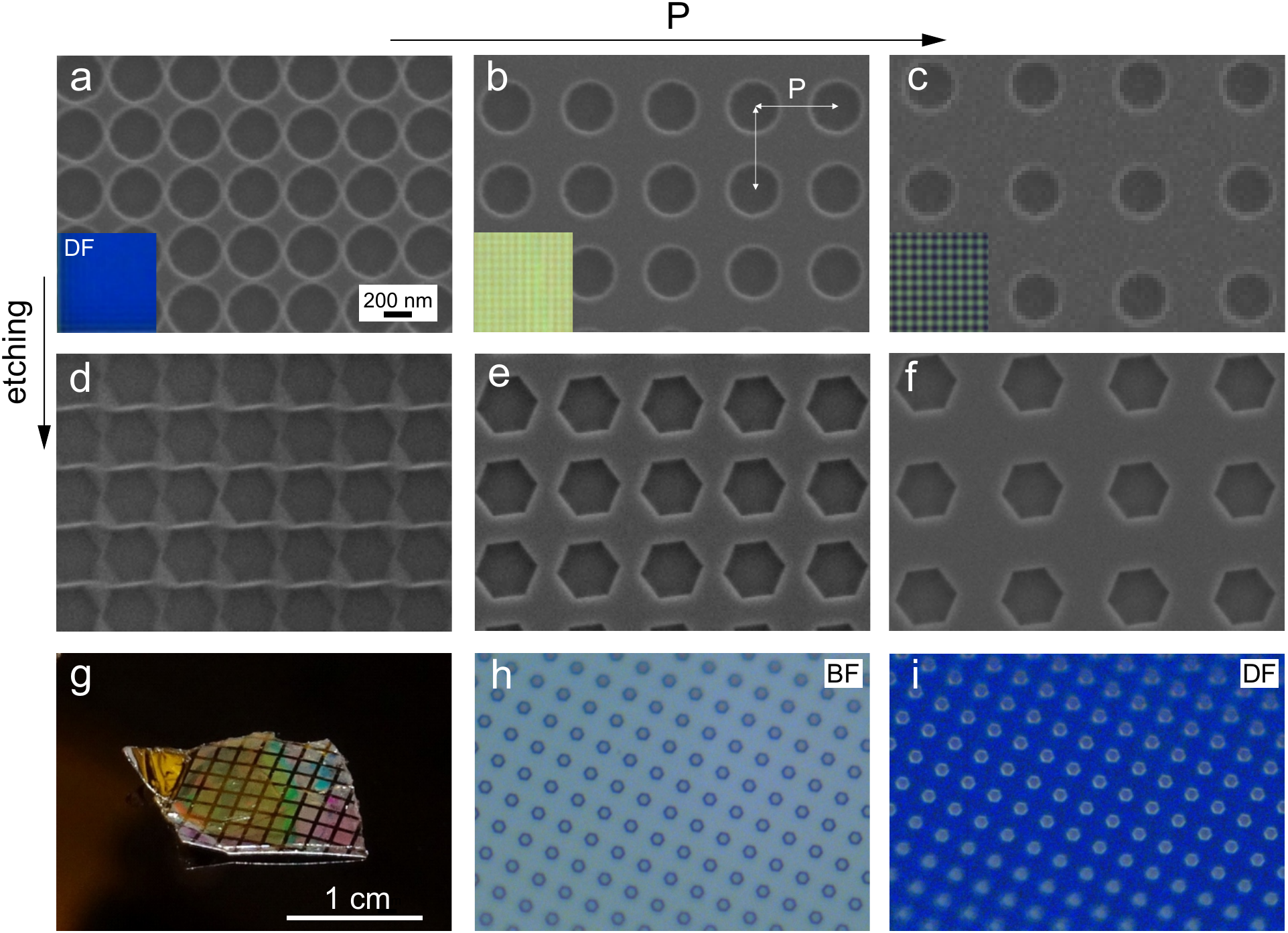}
\caption{\textbf{Arrays of TMD nanoholes.} (a-c) SEM images of nanohole arrays with $D=400$ nm and $P = 100$, 300, and 500 nm before anisotropic wet-etching. Insets show true-color dark-field optical microscope images of the corresponding arrays. (d-e) SEM images of the same nanohole arrays as in (a-c) after anisotropic wet-etching. (g) Optical image of a large-scale (over 1 cm in lateral dimensions) nanopatterned multilayer WS$_2$ crystal. Each colourful patterns in the image is induced by periodically arranged lattices of microholes ($D=1$~$\mu$m) with  $P=2$~$\mu$m. (h) BF and (i) DF images of a hexagonal hole array taken from the large-scale nanopatterned flake.} 
\label{Fig3}
\end{figure}

\subsection{Nanohole arrays}

The next structures we investigated were nanohole arrays in TMD multilayers. Having a possibility to fabricate high quality TMD nanohole arrays of various sizes could open several applications ranging from metamaterials \cite{krishnamoorthy2020infrared} to catalysis \cite{roxlo1987edge, jaramillo2007identification} and sensing \cite{sarkar2014mos2}. Here, we address fabrication of nanohole arrays not only in few-hundred-micron TMD flakes, but also in macroscopic cm-scale TMD crystals. \autoref{Fig3} shows a summary of the obtained WS$_2$ nanohole arrays. Similar to the arrays of nanodisks, the samples were fabricated by a combination of EBL and RIE. The difference was, however, that this was subsequently followed by anisotropic wet-etching to fabricate hexagonal nanohole arrays terminated by zigzag edges \cite{munkhbat2020transition} (Methods). In contrast to the nanodisk arrays, here, we employed a positive resist (ARP 6200.13). First, we fabricated nanohole arrays in mechanically exfoliated WS$_2$ flakes on a transparent and non-conductive glass substrate. The resist (ARP 6200.13) was spin-coated at 3000 rpm, followed by soft baking at 120$^{\circ}$C for 5 min. Subsequently, a thin layer of Cr (20 nm) was deposited directly on top of the resist without any additional e-spacer layer. This simplifies nanopatterning since ARP6200.13 is not modified during Cr removal by the standard Cr etchant (unlike the negative MaN-2405 resist used for nanodisk arrays). TMD flakes are immobilized and protected under the unexposed resist during the development process in $n$-amyl acetate. After development, the sample was etched using a dry RIE process, followed by a few seconds of additional oxygen plasma treatment and removal of the leftover resist in hot acetone.
\autoref{Fig3}a-c shows the fabricated WS$_2$ circular nanohole arrays along with their corresponding DF images. An important difference of nanohole arrays (in comparison to nanodisks) is that they can be selectively wet-etched to produce nearly atomically sharp edges with exclusive zigzag terminations (\autoref{Fig3}d-f) \cite{munkhbat2020transition}. This method was employed to obtain TMDs with sharp zigzag edges after lithography and dry etching processes. Specifically, the sample was exposed to an aqueous solution containing H$_2$O$_2$ and NH$_4$OH (1:1:10 H$_2$O$_2$:NH$_4$OH:H$_2$O volume ratio of stock solutions under mild heating at $T=50^{\circ}$C). This led to formation of well-defined hexagonal holes from pre-patterned circular holes due to anisotropic wet-etching. The resulting zigzag edges could carry a set of useful multi-functional properties, including catalytic activity for hydrogen evolution \cite{jaramillo2007identification, kong2013synthesis} and enhanced charge transport \cite{wu2016uncovering}. Recently, a similar nanofabrication approach using hBN as an etching mask and resulting in nanoholes down to $\sim$20 nm has been introduced \cite{danielsen2021super}.

During anisotropic wet-etching the armchair edges are etched much faster than the zigzag edges, probably because of differences in their chemical reactivity. The resulting nanohole arrays are terminated by sharp zigzag edges, which makes them potentially interesting for applications. For instance, zigzag edges have been predicted to be metallic and ferromagnetic, in contrast to ordinary semiconducting armchair edges \cite{bollinger2001one, li2008mos2, davelou2014mos2}. Zigzag edges could also be useful for high electro- and photo-catalytic activity \cite{zhou2013synthesis}. Moreover, they show potential for nonlinear optics applications, such as enhanced second-harmonic generation \cite{yin2014edge}. In a previous study, we showed that standard WS$_2$ can be tailored with multi-functional atomically sharp zigzag edges using a facile wet-etching technique that uses only abundant and environmentally friendly chemicals \cite{munkhbat2020transition}. Here, we expand this technique to other parameter ranges, improve the nanofabrication procedure, and apply it to macroscopic samples, as we show below. 

In addition to microscopic samples, we fabricated a large-area nanostructured TMD sample in a macroscopic WS$_2$ crystal (size of $\sim 1$ cm in lateral dimensions, see \autoref{Fig3}g-i). The sample was fabricated in a similar manner to the microscopic hole array samples -- by a combination of EBL and RIE, followed by anisotropic wet-etching. To do so, we transferred a few $\mu$m thick WS$_2$ crystal onto a silicon substrate (\autoref{Fig3}g). It is worth mentioning that, in principle, any substrate can be employed or even a bulk TMD crystal can be directly nanopatterned without any additional substrate. \autoref{Fig3}g shows an optical image of a nanopatterned WS$_2$ crystal. Colourful square-patterns on the crystal surface are due to periodically arranged micron-sized holes with a pitch of 2 $\mu$m (see \autoref{Fig3}g). The depth of the etched holes is $\sim$ 3 $\mu$m.  BF and DF images of these hole arrays are shown in \autoref{Fig3}h and \autoref{Fig3}i, respectively.

The advantage of a large-scale nanostructured TMD can be two-fold. First, this allows to prepare nanostructured TMD samples without any resist contamination, by exfoliation directly from the pre-patterned WS$_2$ crystal. We have exfoliated high quality nanostructured TMD samples onto PDMS stamps by employing the standard scotch-tape method from the macroscopic patterned crystal shown in \autoref{Fig3}g. Later, these samples can be transferred onto any desired substrate for further use.
Second, large scale nanopatterned TMDs may be of interest in their own right, since a combination of high surface-to-volume ratio and multi-functional zigzag edges over a large area make them useful for applications in photo- and electro-catalysis \cite{jaramillo2007identification, zhou2013synthesis} and sensing \cite{sarkar2014mos2}. Moreover, the colourful patterns in nanostructured TMD crystals could be useful for structural color and opto-electronic applications. Additionally, this large-scale sample provides evidence that the fabrication techniques developed in this work are potentially scalable and currently limited only by the lateral size of the original TMD crystal.

\section{Nanostructured TMDs for light confinement and waveguiding}

The nanostructures and their applications discussed above are based on a combination of high refractive index, optical anisotropy and high absorption \cite{li2014measurement,wilson1969transition,wang2016direct,munkhbat2018self}. However, TMDs can also support an alternative scenario -- a scenario in which the optical losses are low, while high refractive index and optical anisotropy are preserved \cite{verre2019transition,hu2017probing,li2014measurement,ermolaev2020broadband}. This is realized in the near-infrared spectral range in the case of semiconducting TMDs and can offer novel applications of TMD nanostructures for low-loss nanophotonics, such as high quality factor ($Q$) resonators and waveguides. The combination of low-loss, high refractive index, optical anisotropy and mechanical flexibility \cite{chang2013high} can potentially enrich low-loss nanophotonic applications \cite{ling2021all}.

\begin{figure}
\centering
\includegraphics[width=0.8\textwidth]{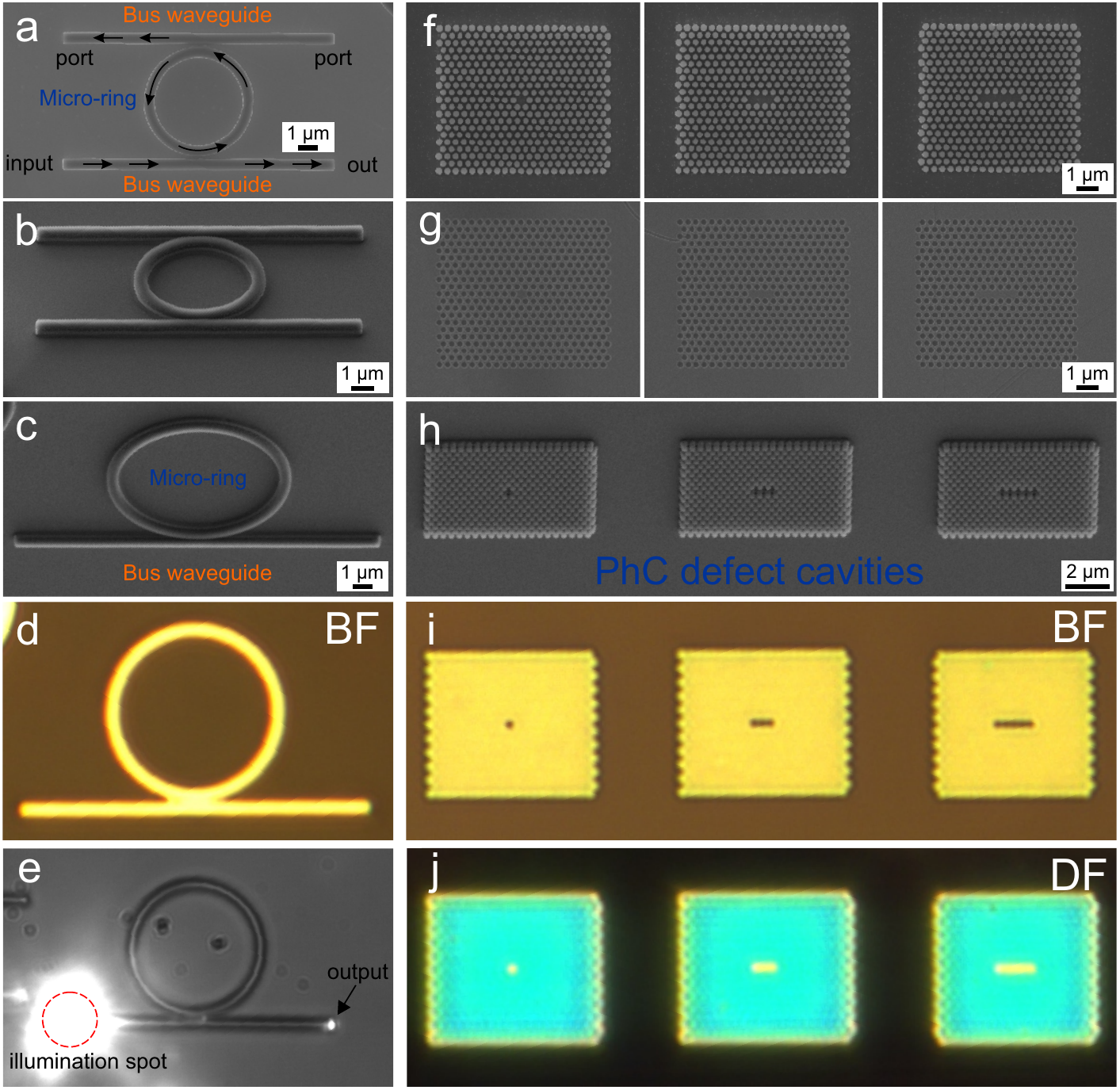}
\caption{\textbf{Exemplary low-loss TMD nanophotonics devices.} (a-e) SEM, BF, and DF images of the fabricated WS$_2$-based nanophotonic circuits consisting of waveguide-coupled WS$_2$ micro-ring resonators.
(f-j) The same as (a-e) for a series of photonic crystal cavities fabricated from a 300 nm thick WS$_2$ film on top of glass.}
\label{Fig4}
\end{figure} 

To make initial steps towards testing these predictions, we have fabricated several proof-of-concept devices (\autoref{Fig4}). Here, we choose multilayer WS$_2$ to fabricate various low-loss nanophotonic devices that can potentially operate at the telecom range (1550 nm). This choice is motivated by the lowest optical loss in WS$_2$ in comparison to other typical TMDs, because the excitonic features (A, B, C excitons) in WS$_2$ appear at lower wavelengths than in other TMDs. \autoref{Fig4}a shows a ring resonator coupled to two waveguides. The sample was fabricated starting from a $L= 150$ nm thick multilayer WS$_2$ on a thermally oxidized silicon substrate with a 3 $\mu$m SiO$_2$ layer and we employed a negative resist (MaN-2405). First, the flake was transferred onto a substrate using a dry transfer technique. The thickness of the flake was measured by a profilometer. Subsequently, theMaN-2405 (500 nm) negative resist was spin-coated for the EBL step. To prevent the adhesion issue of the resist, either the TI:Prime adhesion promoter or a thin layer of SiO$_2$ can be deposited. After the standard EBL process, the resist was developed in the MaD-525 developer to prepare the etching mask. Various designs of microrings and waveguides were fabricated from the multilayer WS$_2$ flake by dry RIE technique (see \autoref{Fig4}a-c for top and tilted-view SEM images). BF and DF images of an exemplary device are shown in \autoref{Fig4}d,e, respectively.

After fabrication we proceeded with testing light coupling into the devices. Specifically, we coupled a broadband white light source (Laser-driven white light source, LDLS) into an input channel of the waveguide using a high numerical aperture objective ($40\times$, $\textrm{NA}=0.95$). As shown in \autoref{Fig4}e, light was transmitted through the waveguide, and the other (output) end of the waveguide was lit up showing a proof-of-concept possibility for TMD-based low-loss photonic devices.

Although we have not carried out optical characterizations of the fabricated devices, we simulated their performance using the finite-difference time-domain (FDTD) method. The simulated data suggest excellent properties of WS$_2$ waveguides and ring resonators (see Figure S8). While the simulated device was not optimized, it shows decent transmission properties with only small reflections at the couplers and very small scattering due to the square meshing of the ring. Our findings are in line with recent calculations suggesting that TMDs could be an excellent low-loss nanophotonics platform \cite{ling2021all}. In our work, we report the first nanofabrication attempts towards realization of this vision.


In addition to the waveguides and ring resonators, we also explored the fabrication of TMD photonic crystal cavities. Perturbing a number of unit cells in a periodic 2D array (for example, removing a series of holes or disks from an array of those) results in the emergence of photonic crystal defect cavities supporting finite lifetime resonances \cite{Molding1995photonic}. Starting from a 300 nm thick WS$_2$ flake on top of a glass substrate hosting a triangular array of holes with a radius of 120 nm and a lattice constant of 400 nm, we fabricated L1, L3, and L5 defect cavities by removing 1, 3, or 5 holes in a row from the periodic array, respectively.
 
\autoref{Fig4}f-j show top and tilted-view SEM images as well as BF and DF images of the fabricated photonic crystal cavities. Since optical characterization of photonic crystal cavities is challenging with far-field spectroscopy and traditionally relies on near-field spectroscopy (or on-chip circuitry), we turned to numerical simulations of the fabricated photonic crystal cavities. Figure S9 presents the resulting electric field distributions of the L1, L3, and L5 cavity modes revealing their fundamental resonances at 1556, 1690, and 1705 nm, respectively. The experimental confirmation of these theoretical predictions is a subject of future work.

\subsection{Nanopatterning of in-plane anisotropic and metallic materials:}

So far we have discussed nanostructuring of standard semiconducting in-plane isotropic TMD materials, such as WS$_2$. However, many TMD materials can be found in other forms, including hyperbolic, and in-plane anisotropic phases. Optical properties of such TMDs are of certain interest and their nanostructuring can open additional means to manipulate light on the nanoscale. Motivated by this, here we developed ways to nanopattern several more exotic TMD materials, including ReS$_2$, TaSe$_2$, TaS$_2$ and NbSe$_2$. These materials possess remarkable properties, for instance, ReS$_2$ is in-plane anisotropic due to its stable 1T'-distorted phase, while  TaSe$_2$, TaS$_2$, and NbSe$_2$ are in-plane isotropic, metallic and even superconducting at low temperatures. Nanopatterning of these more exotic TMD materials can further enrich their properties and open new directions in nanophotonics and nanoelectronics.

\begin{figure}
\centering
\includegraphics[width=1\textwidth]{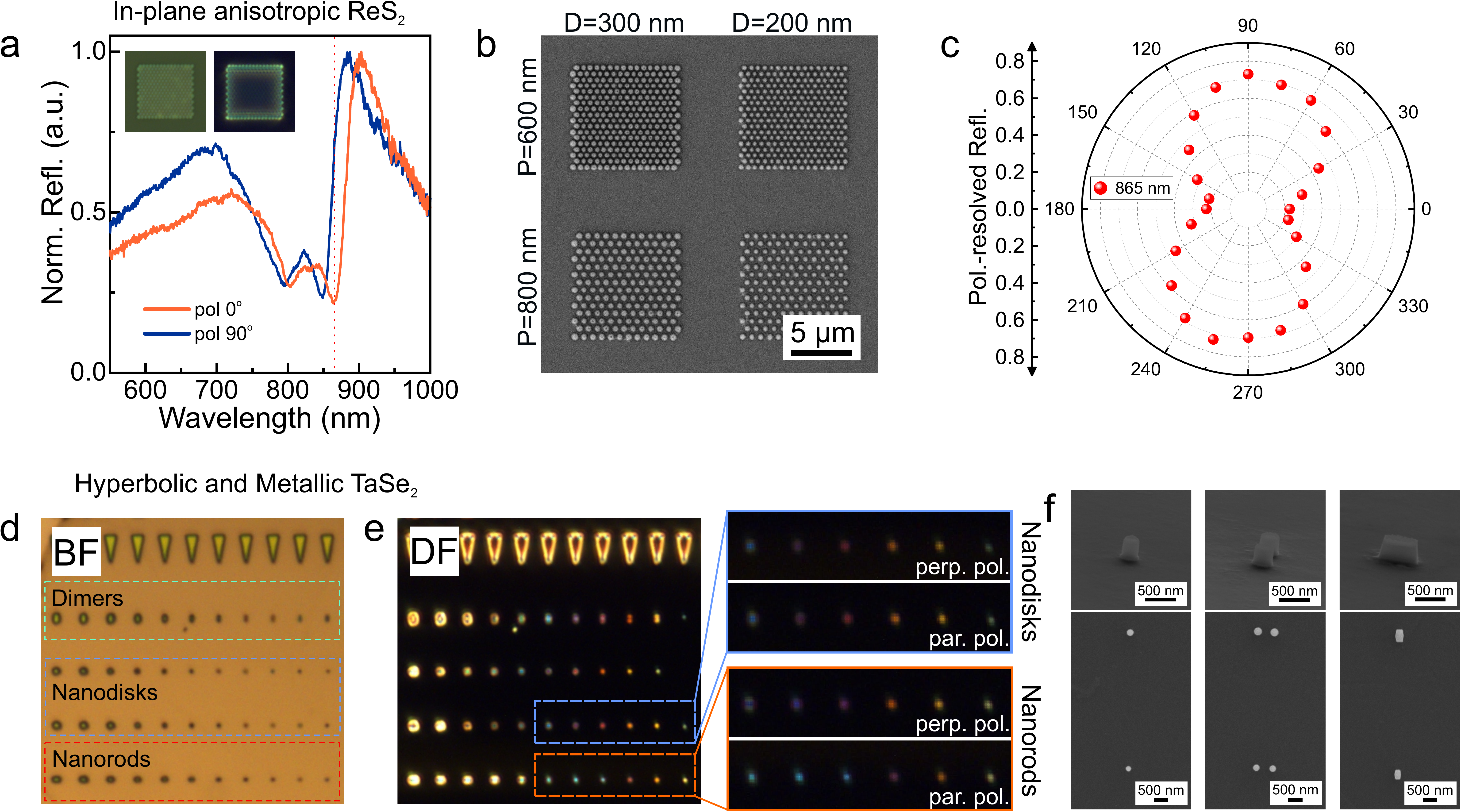}
\caption{\textbf{In-plane anisotropic ReS$_2$.} (a) Reflection spectra at pol-0 (orange) and pol-90 (blue), (b) SEM image, and (c) polarization-resolved reflection shown as a polar plot at 865 nm measured from an exemplary ReS$_2$ nanodisk hexagonal arrays with $D=300$~nm and 200 nm (left and right columns), and $P=600$ nm and 800 nm (top and bottom rows), respectively. Inset in (a) shows DF and BF images of the nanodisk hexagonal array ($D=300$~nm, $P=800$ nm). \textbf{Hyperbolic and metallic TaSe$_2$.} (d) BF and (e) DF images of TaSe$_2$ single nanodisks, nanodisk dimers, and nanorods with various sizes, respectively. Inset shows DF images of nanodisks and nanorods taken under perpendicular and parallel polarization with respect to the longer axis of the nanorod. (f) Corresponding tilted- and top-view SEM images of exemplary TaSe$_2$ single nanodisk, dimer, and nanorod, respectively.}
\label{Fig5}
\end{figure}

We performed several important tests in order to develop a viable nanofabrication strategy for exotic TMDs e.g., ReS$_2$, TaSe$_2$, TaS$_2$ and NbSe$_2$ (see \autoref{Fig5} and Supplementary Figure S10). We started with \ce{ReS2} that is characterized by different in-plane permittivity tensor elements ($\varepsilon_{xx} \neq \varepsilon_{yy}$) and thus enables an in-plane anisotropic optical response. By designing geometrically symmetric nanophotonic structures out of ReS$_2$ with its intrinsic in-plane anisotropy, one could envision photonic crystal structures where polarization-degeneracy would be lifted up. This can be an interesting approach to fabricate e.g. elliptical photonic cavities \cite{wang2019towards}. To demonstrate this possibility in a nanofabrication experiment, we successfully etched nanodisk hexagonal arrays with various dimensions out of 120 nm thick multilayer ReS$_2$ flake using a combination of EBL and dry etching with CHF$_3$ (\autoref{Fig5}a-c). It is important to note that the etching rate of ReS$_2$ with a standard CHF$_3$ gas was about 3$\times$ slower ($\sim$ 3 nm/min) than more conventional TMDs (10 nm/min). This suggests that the etching process for ReS$_2$ is more physical than chemical. The problem is that the selectivity of physical etching is, in general, poor and it also requires relatively thick resists. Therefore, the nanostructuring of thick ReS$_2$ requires further optimization steps and improvements. As a potential improvement, we propose that etching with other chemicals, such as SF$_6$ or Cl-based gases, could make the process more chemical. Another suggestion is to perform etching using hard etching masks such as chromium or aluminum.

\autoref{Fig5}a shows reflection spectra of an exemplary ReS$_2$ nanodisks hexagonal array ($D=300$~nm, $P=800$ nm) with two orthogonal polarizations. As is seen, the polarization dependence of the ReS$_2$ array is anisotropic, due to the in-plane anisotropic crystalline structure of the bulk material. This allows to selectively excite the resonances of the array by tuning the polarization of the light at certain wavelengths.
An exemplary SEM image of the ReS$_2$ array is shown in \autoref{Fig5}b, which confirms our nanofabrication strategy is successful and applicable to ReS$_2$. Furthermore, the polar plot in \autoref{Fig5}c shows polarization-resolved reflection at 865 nm (close to maximum anisotropy), demonstrating pronounced optical anisotropy inherited from the bulk crystal. Such anisotropy implies that nanostructured multilayer ReS$_2$ could be useful for SHG applications due to its natural broken inversion symmetry.

Furthermore, we fabricated TaSe$_2$ nanodisks and nanorods out of 150 nm freshly exfoliated TaSe$_2$ flake. This material has a metallic response for in-plane electronic excitations (with the plasma frequency in the near infrared range), while a dielectric response for out-of-plane electronic excitations, which means the material is naturally hyperbolic. BF, DF and SEM images of the prepared samples are shown in \autoref{Fig5}d,e. TaSe$_2$ nanodisks were fabricated with various diameters ranging from 100 nm to 1000 nm, whereas nanorods were elongated with the aspect ratio ($AR$) of 2 along one of the axes. In \autoref{Fig5}e, one can clearly observe size- and shape-dependent colors in the DF image, in line with a typical plasmonic and dielectric nanophotonics size dispersion.
Polarization-resolved DF images in the inset of \autoref{Fig5}e were obtained under perpendicular and parallel polarization with respect to the long axis of the nanorod. Polarization-resolved DF images from TaSe$_2$ nanodisks ($D$: 100 nm - 300 nm) exhibit identical symmetric response under different polarizations. On the other hand, elongated TaSe$_2$ nanorods ($AR$=2) show polarization dependent response due to asymmetric geometry. Since the plasma frequency of TaSe$_2$ and similar metallic TMDs occurs in the near-infrared region around $\sim$1000 nm - 1300 nm \cite{gjerding2017band}, a successful nanostructuring of such hyperbolic TMDs could lead to a combined 2-in-1 plasmonic-high-index-dielectric response in the visible -- near-infrared range, with plasmonic response dominating at frequencies below the plasma frequency, while dielectric response -- at frequencies above the plasma. Our nanofabrication strategy successfully demonstrates the feasibility of this approach.

\section{Conclusion}
In conclusion, in this work we propose a roadmap of all-TMD nanophotonics for light absorption and light confinement and guiding. We demonstrate several viable nanofabrication strategies towards reaching key nanophotonic building blocks, including nanodisk and nanohole arrays, ring resonators, waveguides and photonic crystals on several relevant substrates. TMD nanophotonics may take advantage of TMD's remarkable intrinsic optical properties, namely, high refractive index ($n \sim$ 4), low optical loss (in the near-infrared region), in-plane and out-of-plane anisotropy ($\varepsilon_{xx}=\varepsilon_{yy}\neq\varepsilon_{zz}$ and $\varepsilon_{xx}\neq\varepsilon_{yy}\neq\varepsilon_{zz}\neq\varepsilon_{xx}$, respectively),
and hyperbolicity ($\varepsilon_{xx}=\varepsilon_{yy}<0$ and $\varepsilon_{zz}>0$). Special attention was devoted to nanofabrication on non-conductive transparent substrates (SiO$_2$), which are commonly used in nanophotonic applications. The nanofabrication methods developed in this work are desperately needed to open new possibilities for TMD nanophotonics \cite{ling2021all}. We envision that nanostructured TMDs will enrich the toolbox of available materials for light absorption, waveguiding, metasurfaces, and other applications.

\section{Acknowledgements}
B.M., D.G.B, B.K., and T.O.S. acknowledge financial support from the Swedish Research Council (VR Miljö project, grant No: 2016-06059), the Knut and Alice Wallenberg Foundation (grant No: 2019.0140), Chalmers Excellence Initiative Nano and 2D-TECH VINNOVA competence center (Ref. 2019-00068). D.G.B. acknowledges Council on grants of the President of the Russian Federation (MK-1211.2021.1.2) and Russian Science Support Foundation (21-72-00051). T.J.A. thanks the Polish National Science Center for support via the projects 2019/34/E/ST3/00359 and 2019/35/B/ST5/02477.

\bibliographystyle{apsrev4-1}

\bibliography{TMDCzoo}

\end{document}